\documentclass[pdflatex,sn-nature,iicol]{sn-jnl}

\usepackage{threeparttablex}%
\usepackage{graphicx}%
\usepackage{multirow}%
\usepackage{amsmath,amssymb,amsfonts}%
\usepackage{amsthm}%
\usepackage{mathrsfs}%

\usepackage[title]{appendix}%
\usepackage{xcolor}%
\usepackage{textcomp}%
\usepackage{manyfoot}%
\usepackage{booktabs}%
\usepackage{algorithm}%
\usepackage{algorithmicx}%
\usepackage{algpseudocode}%
\usepackage{listings}%
\usepackage{siunitx}%
\usepackage{tikz} 
\usetikzlibrary{positioning,arrows.meta}

\begin{document}

\title[A pathway towards decentralized studies of radioactive post-lead elements and their applications in beyond standard model physics]{A pathway towards decentralized studies of radioactive post-lead elements and their applications in beyond standard model physics}


\author*[1]{\fnm{Moritz Pascal} \sur{Reiter}}\email{mreiter@ed.ac.uk}

\author[2,3]{\fnm{Kriti} \sur{Mahajan}}
\author[4,5]{\fnm{Meetika} \sur{Narang}}
\author[6]{\fnm{Carsten} \sur{Z\"ulch}}
\author[2,4]{\fnm{Timo} \sur{Dickel}}

\author[2,4]{\fnm{Daler} \sur{Amanbayev}}
\author[6]{\fnm{Robert} \sur{Berger}}
\author[2]{\fnm{Julian} \sur{Bergmann}}
\author[1]{\fnm{Agnieszka} \sur{Bukowicka}}
\author[6]{\fnm{Mariam} \sur{Fadel}}
\author[1,4]{\fnm{Tayemar} \sur{Fowler-Davies}}
\author[4]{\fnm{Zhuang} \sur{Ge}}
\author[4]{\fnm{Simeon} \sur{Gloeckner}}
\author[1,2]{\fnm{Gabriella} \sur{Kripk\'o-Koncz}}
\author[5]{\fnm{Nasser} \sur{Kalantar-Nayestanaki}}

\author[1]{\fnm{Cameron} \sur{Merron}}
\author[4,7]{\fnm{David J.} \sur{Morrissey}}
\author[2,4]{\fnm{Wolfgang R.} \sur{Pla{\ss}}}
\author[2,3,4]{\fnm{Christoph} \sur{Scheidenberger}}
\author[2,3]{\fnm{Makar} \sur{Simonov}}
\author[4,8]{\fnm{Nazarena} \sur{Tortorelli}}
\author[4]{\fnm{Jiajun} \sur{Yu}}
\author[1,2]{\fnm{Alexandra} \sur{Zadvornaya}}
\author[4]{\fnm{Jianwei} \sur{Zhao}}

\affil[1]{\orgdiv{School of Physics and Astronomy}, \orgname{University of Edinburgh}, \orgaddress{\city{Edinburgh}, \postcode{EH9 3FD}, \country{United Kingdom}}}

\affil[2]{\orgdiv{II. Physikalisches Institut}, \orgname{Justus-Liebig Universitaet Giessen}, \orgaddress{\city{Giessen}, \postcode{35392}, \country{Germany}}}

\affil[3]{\orgdiv{Helmholtz Research Academy Hesse for FAIR (HFHF)}, \orgname{GSI Helmholtz Center for Heavy Ion Research, Campus Gießen}, \orgaddress{\city{Giessen}, \postcode{35392}, \country{Germany}}}

\affil[4]{\orgname{GSI Helmholtz Centre for Heavy Ion Research GmbH}, \orgaddress{\city{Darmstadt}, \postcode{64278}, \country{Germany}}}

\affil[5]{\orgdiv{Nuclear Energy Group ESRIG}, \orgname{University of Groningen}, \orgaddress{\city{Groningen}, \postcode{9747 AA}, \country{The Netherlands}}}

\affil[6]{\orgdiv{Department of Chemistry}, \orgname{Philipps-Universit\"at Marburg}, \orgaddress{\city{Marburg}, \postcode{35032}, \country{Germany}}}

\affil[7]{\orgdiv{Department of Chemistry}, \orgname{Michigan State University}, \orgaddress{\city{East Lansing}, \postcode{MI 48824}, \country{United States}}}

\affil[8]{\orgdiv{Faculty of Physics}, \orgname{Ludwig-Maximilians-Universit{\"a}t M{\"u}nchen}, \orgaddress{\city{Garching}, \postcode{85748}, \country{Germany}}}


\abstract{
Molecules have proven to be sensitive tools for studying physics beyond the standard model, with heavy and deformed nuclei offering decisive sensitivity to parity- and time-reversal-violating effects. However, almost all elements beyond lead, occupying the 6p~to~5f atomic orbitals, lack stable isotopes, hence molecules containing them are referred to as radioactive molecules. Among those, radium monofluoride has seen particular interest, but to date, research on radioactive molecules has mainly been limited to large-scale nuclear facilities. 
Here, we present a scheme that allows efficient and fast harvest of radioactive ions (including short-lived Ra), and show ion gas-phase reaction studies of singly and doubly charged Ra, Po, and Pb ions with SF$_6$ gas inside an ion trap. 
 Our proof-of-concept study show that the chemical reaction rate of Ra$^+$ is in line with trends of other alkaline earth elements, further supported by quantum chemical computations. 
The reaction Ra$^{2+}$ + SF$_6$ $\rightarrow$ RaF${^+}$ + SF$_5^{+}$ achieves an almost unity conversion efficiency, making it particularly suitable for the application for studies in physics beyond the standard model. 
The scheme enables future decentralized research avenues with short-lived radioactive molecules for fundamental physics research at laboratories without the need for local nuclear reactors or accelerators.

}






\maketitle

\section{Introduction}\label{sec1} 

Recent searches for physics beyond the Standard Model (SM) have been motivated by observations such as the baryon asymmetry of the Universe \cite{RevModPhys.76.1}, the existence of dark matter \cite{RevModPhys.90.045002}, and the acceleration of cosmic expansion \cite{Perlmutter_1999}, none of which are fully accounted for within the SM. Precision tests of a violation of the combined symmetries charge, conjugation, and parity (CP) via permanent electric dipole moments (EDMs) of fundamental particles provide a particularly sensitive and model-agnostic pathway to new physics \cite{PhysRevLett.19.1396}. Over the past decade, atomic \cite{PhysRevLett.124.081803,PhysRevLett.114.233002,PhysRevLett.88.071805,graner2016} and molecular \cite{Andreev2018,doi:10.1126/science.adg4084} experiments have delivered constraints competitive with, and complementary to, high-energy collider searches \cite{Abada2019}.

Heavy, polar molecules are especially powerful platforms for EDM searches \cite{demille:2015}. Their internal structure enables strong polarization in modest laboratory fields \cite{PhysRevLett.19.1396}, closely spaced opposite-parity levels enhance sensitivity to parity and time-reversal violating effects \cite{kozlov1995,PhysRevLett.100.023003}, and effective internal fields increase rapidly with nuclear charge $Z$, with enhancements often scaling with $\sim Z^3$ \cite{RevModPhys.90.025008,gaul2024}. Additional nuclear structure features in post-lead elements (e.g. Fr, Ra, Ac, Pa, Th, \dots) - in particular large quadrupole and octupole deformations - can further amplify sensitivity to hadronic CP violation \cite{PhysRevC.93.044304,PhysRevC.100.044321,RevModPhys.91.015001}. Quantum chemistry and nuclear theory have thus identified a family of promising fluoride, oxide, hydroxide, and nitride species featuring heavy elements \cite{gaul2024,PhysRevA.82.052521,PhysRevA.101.042504,gaul2019}.
Despite this promise, experimental progress continues to be constrained by isotope availability. All elements beyond bismuth are radioactive; only a few isotopes exist in macroscopic quantities, while many of the most attractive candidates are short-lived and must be generated at online radioactive ion beam (RIB) facilities or research reactors, with great efforts and in minute amounts only \cite{Arrowsmith-Kron_2024}. Even if microgram quantities can be produced, such routes typically require large facilities, can be chemically challenging for poorly known elements, and are intrinsically too slow for short-lived isotopes. This practical bottleneck limits systematic surveys of post-lead and actinide molecules. To date, landmark EDM measurements have focused molecules with naturally occurring isotopes such as ThO \cite{Andreev2018}, HfF$^+$ \cite{PhysRevLett.119.153001}, and ThF$^+$ \cite{ng:2022}, 
while only a few artificially produced radioactive molecular species (RaF and AcF) have been studied in detail \cite{GarciaRuiz2020,Athanasakis-Kaklamanakis2025_2,10.1063/5.0159888}. 


Beyond fundamental symmetry tests, access to high-$Z$ isotopes is also central to radiochemical and medical applications. They allow tests of periodic trends under extreme relativistic effects, explore atomic and chemical properties at the limits of the periodic table \cite{https://doi.org/10.1002/anie.200685471} or guide the development of radio-pharmaceuticals; for example At and Ac are leading candidates for targeted $\alpha$-therapy \cite{10.3389/fmed.2022.1076210,10.3389/fmed.2022.1030094}. Yet their basic chemistry remains poorly characterized primarily due to their accessibility and short half-lives \cite{https://doi.org/10.1002/anie.200685471}. 

Ion–molecule reactions of cations with neutrals benchmark capture dynamics and long-range forces, reveal bond energies, and provide rate and thermochemical data. Extensive bodies of work span much of the periodic table, mapping reactivity with small molecules (e.g., O$_2$, H$_2$O, CO, NH$_3$) \cite{anicich2003index,https://doi.org/10.1002/mas.21703}, including reactions with SF$_6$ \cite{10.1021/jp808077b,10.1021/ic061000o}. Moreover, different charge states can substantially alter reaction pathways, kinetics, and product distributions \cite{10.1021/jp0307194,BOHME2022116674}.
However, the gas-phase ion chemistry of post-lead elements remains sparsely explored \cite{https://doi.org/10.1002/anie.200685471}. 
Efforts to supply longer-lived post-lead elements (half-lives $\gtrsim 10$~days) and molecules are underway at many research laboratories \cite{PEN201462,AU2023375,BALLOF2023224}
and parallel work using microgram quantities via chemical separation and ionization has enabled important studies on Th, U, Np, Pu, and Am, including ion gas-phase reactions with singly-charged M$^+$ \cite{10.1021/acs.jpca.2c02090} and double-charged M$^{2+}$ \cite{10.1021/jp0447340}, 
but to date, work has mainly been limited to long-lived isotopes.

 Several complementary routes toward laboratory-scale studies of short-lived atomic radium and radium-bearing molecules have emerged recently. Thorium-generator-based in-vacuo sources have enabled photoionization loading, trapping, and laser cooling of $ ^{224} $Ra$ ^+ $ and $ ^{225} $Ra$ ^+ $ ions for precision spectroscopy and quantum control \cite{Ready2024arxiv,Fan2023PRR}. In parallel, cryogenic buffer-gas methods have enabled the production, cooling, and high-resolution laser spectroscopy of neutral radium-containing molecules, including RaOH and RaF, in a tabletop setting \cite{Conn2025arxiv}.

 Here, we demonstrate a broadly applicable and accelerator-independent scheme, complementary to existing approaches, to produce and identify short-lived heavy radioactive molecular ions at trace level sensitivity.
Our method combines (i) alpha-recoil harvesting in a high-purity gas-filled stopping cell with (ii) preparation of a clean ion sample, (iii) fast and selective ion–molecule reactions in an ion trap, and (iv) high-resolution and broadband identification by mass spectrometry. Steps (ii) and (iii) are done inside a versatile linear Radio Frequency Quadrupole (RFQ) beamline. We generate and detect RaF$^+$, PoF$^+$, and PbF$^+$, quantify their reaction rates, and establish charge-state selectivity. We show that M$^{2+}$ ions, with M = $^{224}$Ra (t$_{1/2}=3.6$~days), $^{216}$Po (t$_{1/2}=145$~ms), $^{212}$Pb (t$_{1/2}=10.6$~hours), react efficiently with SF$_6$ to form monofluorides (MF$^+$), while M$^+$ pathways are strongly element dependent.
State-of-the-art quantum chemistry calculations support the observed trends by predicting exothermic formation of MF$^+$ in the M$^{2+}$ + SF$_6$ gas-phase reaction and the observed suppression of endothermic M$^{+}$ channels.

The production method addresses key bottlenecks for decentralized studies of molecules bearing post-lead elements: (i) it decouples molecular production from slow wet-chemistry separations, (ii) operates at single ion sensitivity levels (atom at a time) suitable for university laboratories, (iii) proceeds on millisecond time scales compatible with short half-lives, and (iv) provides high chemical specificity via mass-resolved detection. 

The present work focuses on RaF due to its importance in P- and CP-violation studies \cite{PhysRevA.82.052521}, 
but the technique can be generally applied to a broad set of isotopes and molecules beyond $^{208}$Pb, enabling systematic studies relevant to Beyond Standard Model searches, heavy-element chemistry, and radionuclide applications.

\section{Results}\label{results}

\begin{figure*}[hbt]
    \centering
    \includegraphics[width=0.9\linewidth]{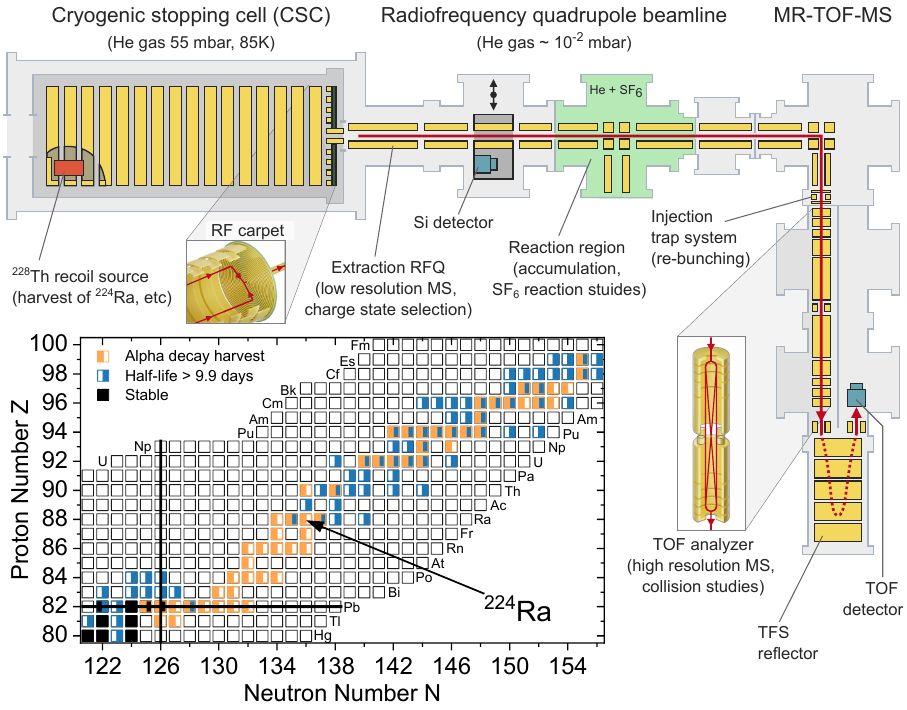}
    \caption{\textbf{Overview of the recoil-harvesting and molecular-ion identification scheme} Schematic of the experimental setup at the FRS Ion Catcher, showing the cryogenic stopping cell (CSC), the versatile radio-frequency quadrupole beam line, and the high-resolution multiple-reflection time-of-flight mass-spectrometer (MR-TOF-MS). The setup has an overall length of about 4~m. Inset, section of the nuclear chart showing isotopes beyond $^{208}$Pb which can be made available for studies at university laboratories via the in-stopping cell alpha decay harvest scheme using readily available parents in comparison to long-lived (half-life $> 9.9$~days) isotopes. The elements At, Rn and Fr do not have any long-lived isotope and as such they are almost impossible to be studied via conventional wet chemistry. (Note, using online production, additional parent isotopes can be produced to further expand the harvesting scheme.)
    }
    \label{fig:setup}
\end{figure*}

\normalsize
Gas-filled stopping cells \cite{WADA2013450}, devices filled with noble gases like helium or argon, are commonly used in nuclear physics research to slow down and thermalize fast ions ($\geq$MeV kinetic energy)  
for low-energy precision experiments (eV-keV kinetic energy). These devices are found at radioactive beam facilities such as IGISOL \cite{AYSTO2001477}, GSI/FAIR \cite{Purushothaman_2013}, 
RIKEN \cite{SCHURY2016425}, FRIB \cite{LUND2020378}, and others to stop fast reaction products, and also can be used to thermalize and extract spontaneous fission fragments 
\cite{SAVARD20084086}. 
Modern,  cryogenic, stopping cells reach excellent purity levels (noble gas purity ppb or better) and can achieve collection efficiencies of approximately $50$--$80\%$ \cite{Purushothaman_2013}. Under such ultra-clean conditions, the ions of most elements are extracted as singly, doubly, or even triply charged species. 
After extraction, the ions can be delivered to subsequent experiments as high-quality, low-emittance ion beams \cite{WADA2013450}, thus providing beams for the investigation of singly and multiply charged ion-chemistry.

Decay harvest schemes, which among others have been used for studies towards a thorium-based nuclear clock \cite{v.d.Wense2015}
or laser studies of Ra$^+$ and RaOH$^+$ \cite{Ready2024arxiv,PhysRevLett.126.023002,Fan2023PRR,Conn2025arxiv}, can be expanded to the whole region of isotopes beyond $^{208}$Pb, see Fig.~\ref{fig:setup} showing potentially accessible nuclides.
Stopping cells are an ideal tool to harvest nuclear decay products following alpha decay. The recoil energy is sufficiently small, approximately $100$~keV in the case of alpha decay of heavy elements, that the ions can be stopped and thermalized in less than a cm of buffer gas  ($50$~mbar/$300$~K), while reaching high extraction efficiencies.
 Thus, compact stopping cells of order $ 10^3 $~cm$^3 $, operated at moderate helium pressures and even at room temperature, can stop and extract the corresponding $\alpha$ recoils, with fast extraction times ($<5$~ms) reducing the cleanliness requirements.

The combination of a stopping cell with an ion trap integrated in a versatile and modular RFQ-based beamline can be used to create specific molecular ions, which can then be identified and quantified using a high-resolution Multiple-Reflection Time-of-Flight Mass Spectrometer (MR-TOF-MS). In-trap ion-gas phase reactions, following similar concepts as studies on naturally occurring elements \cite{10.1021/ic061000o,10.1021/jp0307194}, are particularly suitable for radioactive isotopes due to the high selectivity and sensitivity to trace amounts of material, inherent to experiments with radioactive isotopes \cite{AU2023375}. 
Thus, this approach offers an efficient and practical way to provide a wide range of isotopes of elements beyond lead and molecules containing those for fundamental research. 
It enables decentralized studies across multiple fields using well-controlled small quantities of post-lead elements. 

Among many molecules, radium monofluoride, RaF, has emerged as a potent candidate for physics beyond standard model searches \cite{PhysRevA.82.052521,PhysRevA.90.052513} and has received high attention with experimental work, so far, almost exclusively performed at the online radioactive beam facility ISOLDE, CERN, \cite{GarciaRuiz2020,Athanasakis-Kaklamanakis2025_2,wilkins2025}.  
Here, we successfully perform the first ion-gas phase reaction studies of singly and doubly charged radioactive Ra, Po, and Pb ions with SF$_6$ gas using the FRS Ion Catcher facility \cite{PLA2013457} at the GSI research center, Germany. The work was performed without using the GSI accelerators and as such is widely applicable at local university laboratories.

\subsection{Observation and Identification of Radioactive Molecules}\label{observation}

The experiment was performed at the research facility GSI Helmholtzzentrum für Schwerionenforschung, Darmstadt, Germany, using the FRS Ion Catcher \cite{PLA2013457}. The experimental setup, shown schematically in Fig.~\ref{fig:setup} (top), consists of a cryogenic stopping cell (CSC) \cite{Purushothaman_2013} coupled to a versatile Radio-Frequency Quadrupole (RFQ) beam line \cite{GREINER2020324} and high-resolution MR-TOF-MS \cite{DICKEL2015172}. A commercial $^{228}$Th source (Eckert and Ziegler)  with an activity of about $ 7 $~kBq was installed inside the CSC. In the cryogenic cell, filled with 55~mbar He gas, alpha decay recoils are stopped within a few mm from the source surface. 
Once thermalized in the buffer gas, a combination of DC and RF electric fields was used to extract the recoil ions into the differentially pumped RFQ beam line. Ra, Po, and Pb ions were predominantly extracted in their $2+$ charge state, whereas Rn and Tl were extracted as singly charged ions. 
In the RFQ beam line, operated at a residual gas pressure of approximately $ 10^{-2}$~mbar He, ions were continuously cooled via collisions with the room temperature He buffer gas.
As indicated in Fig.~\ref{fig:setup}, one vacuum chamber was used as the reaction region, where SF$_6$ gas could be mixed with the He buffer gas.
Following this reaction region, precursor and/or product ions could then be transported to an injection trap system, where they were re-bunched and then injected into the time-of-flight analyzer of the MR-TOF-MS. 
In our experiment, the mass spectrometer was operated in three different modes a) ions underwent no reflections to monitor a wide mass range as possible with a mass resolving power m$/\Delta$m of $~2\ 000$ FWHM, b) an intermediate mode, where ions underwent 9 turns allowing unambiguous identification in a mass window from $209$ to $253$ AMU with a mass resolving power of m$/\Delta$m of $~20\ 000$ FWHM and c) a high-resolution mode, where ions were stored for 100 turns inside the mass analyzer resulting in a mass resolving power m$/\Delta$m of $~100\ 000$ FWHM.  Detected rates at the MR-TOF-MS were about $10$~ions~s$ ^{-1} $.
Adding SF$_6$ gas at a partial pressure of  $P_{SF_6}=9(6)\cdot 10^{-5}$~mbar to the He gas  ($P_{He}=9(4) \cdot 10^{-3}$~mbar) in the reaction region allowed gas phase ion-molecule reactions during transport, where ions were kept for a minimum of $180~\mu$s in the SF$_6$ region. Under these conditions, small amounts of RaF${^+}$ and SF$_5^{+}$ ions could be identified in the mass spectrum, demonstrating that the reaction of Ra$^{2+}$ + SF$_6$ $\rightarrow$ RaF${^+}$ + SF$_5^{+}$ was occurring.  

\begin{figure}[htb]
    \centering
    \includegraphics[width=0.8\linewidth]{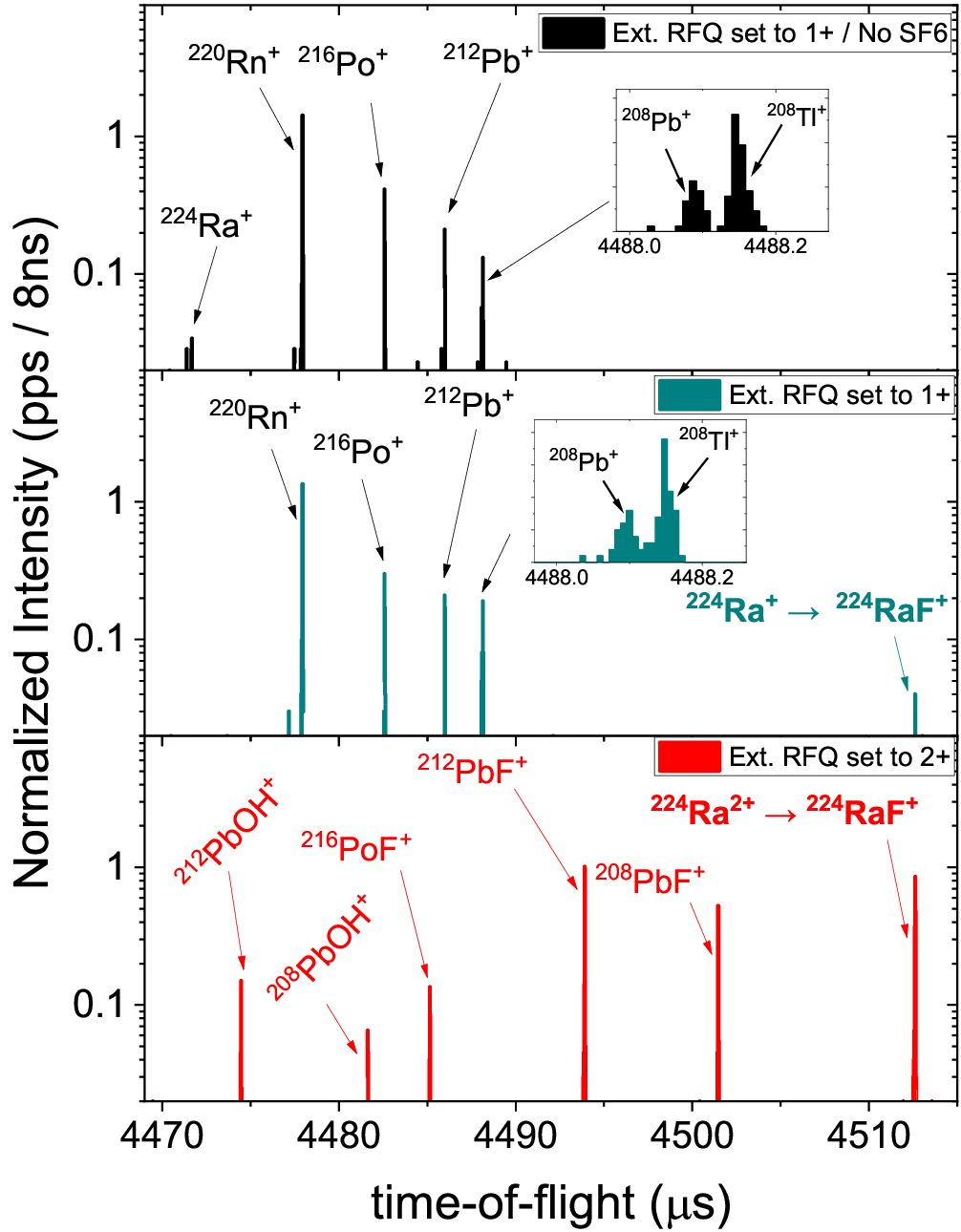}
    \caption{\textbf{Charge-state-selective fluorination identified by MR-TOF mass spectrometry} High-resolution time-of-flight spectrum showing the observed singly charged species without SF$_6$ present in the reaction region (top) and observed reaction products from singly (mid) and doubly (bottom) charged atomic ions with SF$_6$. Note: RaF$^+$ is formed from reactions of both charge states, whereas PoF$^+$, PbF$^+$, and PbOH$^+$ are only formed from $2+$ ions. No reactions of Tl were observed.}
    \label{fig:mass_spec}
\end{figure}

To confirm the identification of the individual ion species, the MR-TOF-MS was switched into high-resolution mode. The number of turns was chosen such that mass $220$~u ions underwent $101$ turns inside the TOF analyzer, resulting in a high-resolution mass spectrum covering the mass-over-charge range of $206 - 244$~u/$e$ ions. Based on their respective time-of-flight, the previous identification of Ra$^{2+}$, RaF${^+}$ and SF$_5^{+}$ were confirmed with high confidence, see Fig.~\ref{fig:mass_spec} for a set of example time-of-flight spectra. 
In the high-resolution mass spectrum additional fluoride molecules $^{208,212}$PbF$^{+}$ and $^{216}$PoF$^{+}$ as well as hydroxides $^{208,212}$PbOH$^{+}$ were unambiguously identified based on their known mass-over-charge following procedures as in \cite{PhysRevC.99.064313}. 


\subsection{Reaction study of radium with SF$_{6}$}
\begin{figure}[htb]
    \centering
    \includegraphics[width=0.8\linewidth]{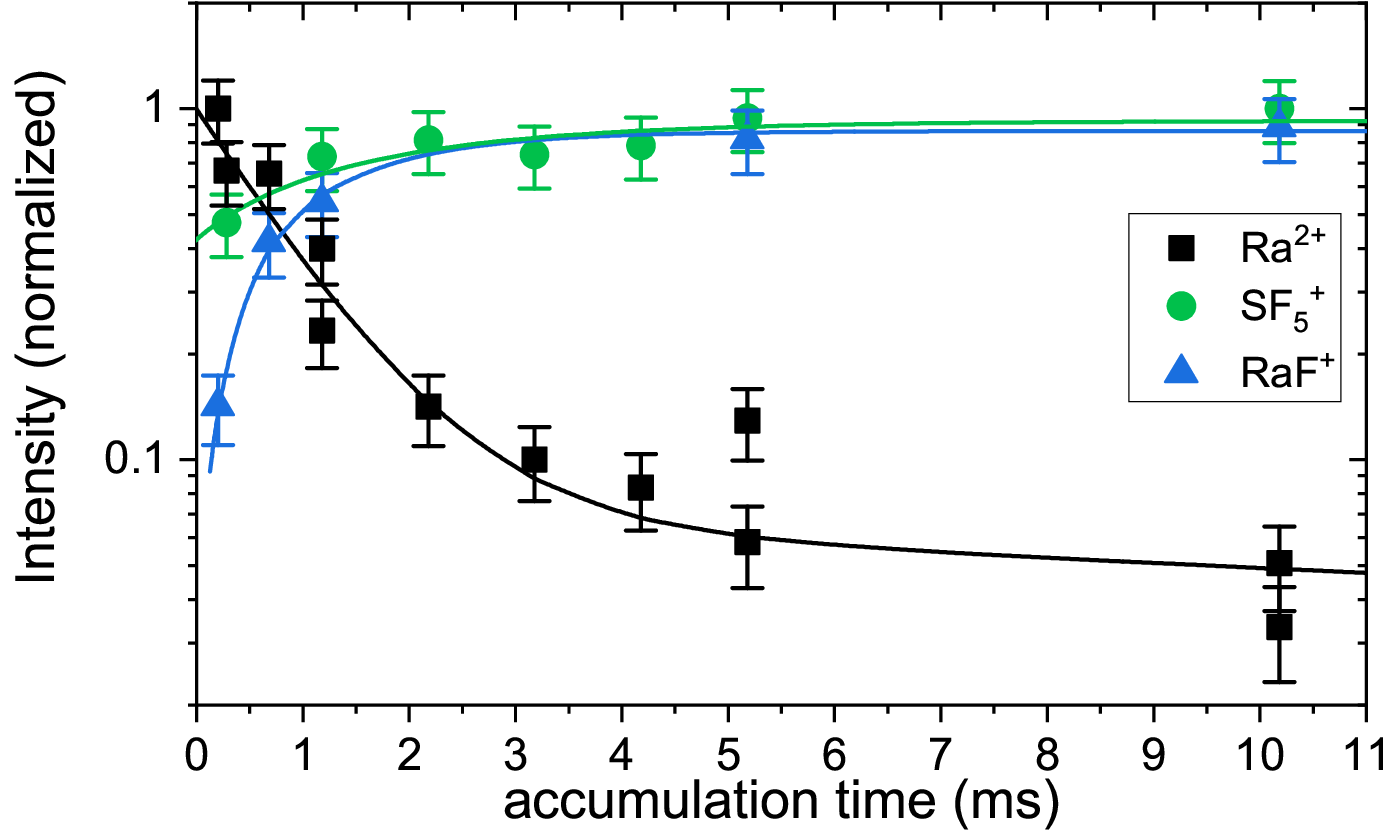}
    \caption{\textbf{Rapid conversion of $^{224}$Ra$^{2+}\rightarrow ^{224}$RaF$^{+}$} Relative intensities of $^{224}$Ra$^{2+}$, $^{224}$RaF$^{+}$ and SF$_5^+$ as a function of accumulation time, normalized to initial $^{224}$Ra$^{2+}$ intensity; demonstrating the reaction $^{224}$Ra$^{2+} + $SF$_6 \rightarrow ^{224}$RaF$^{+} + $SF$_5^+$ using the broad band mode of the MR-TOF-MS. }
    \label{fig:Radium_decay}
\end{figure}

To study the Ra$^{2+}$ + SF$_6$ $\rightarrow$ RaF${^+}$ + SF$_5^{+}$ gas-phase reaction in more detail, the electric fields in the reaction region were adjusted to form an ion trap.  
First, it was shown that Ra$^{2+}$ ions could be stored with minimal losses for extended accumulation times,  confirming the cleanliness and suitability of the environment for controlled ion--molecule studies. The decay constant for ions in the trap was found to be $\tau_{\mathrm{Ra}^{2+}}$(No~SF$_6$)$=900(300)$~ms.
Then, SF$_6$ gas was reintroduced, and radium ions were accumulated and stored in the reaction region. 
The intensities of Ra$^{2+}$, RaF${^+}$, and SF$_5^{+}$ were measured over a range of different accumulation times, as shown in Fig.~\ref{fig:Radium_decay}. 
Note that most of the initial Ra$^{2+}$ ions were converted into RaF${^+}$ ions after an accumulation time of only 10~ms. Our data indicates a conversion efficiency of $90(10)\%$ for producing RaF${^+}$ ions via the Ra$^{2+}$ + SF$_6$ reaction. 

The results from the Ra$^{2+}$ + SF$_6$ $\rightarrow$ RaF${^+}$ + SF$_5^{+}$ reaction are well described by an effective first-order reaction mechanism. 
Thus, the reduction in Ra$^{2+}$ signal was fitted  using a least-squares minimization with an exponential decay curve $N(t)=N_0 \cdot \mathrm{e}^{-t/\tau}$ resulting in a time constant of $\tau_{\mathrm{Ra}^{2+}}=1(0.3)$~ms, whereas the growth of both RaF${^+}$ and SF$_5^{+}$ follow a growth curve $N(t)=N_{0} (1 - \mathrm{e}^{-t/\tau})$ with time constants of $\tau_{\mathrm{RaF}^{+}}=1.2 (0.3)$~ms and $\tau_{\mathrm{SF}_5^{+}}=2.3 (0.9)$~ms, respectively. 
Ra$^{2+}$ decay and RaF${^+}$ growth are in good agreement with each other. The growth of SF$_5^{+}$ seems to be about a factor of two slower, but as the SF$_5^{+}$ intensity had already reached about $50\%$ of its saturation intensity at the lowest accumulation time, the time constants carries a large fit uncertainty. Overall, all three are in agreement within $1.4~\sigma$.   

\subsection{Charge state selectivity during M$^{n+}$+SF\textsubscript{6} reactions}

It is well known that large changes in chemical reactivity may occur with changes in the charge state of the reactant ion \cite{BOHME2022116674}. 
The extraction RFQ in the RFQ beam line was used as a low-resolution quadrupole mass filter to select the charge state delivered to the reaction region, while keeping the MR-TOF-MS monitoring mass/charge $205 - 245$~u/$e$ ions at high resolution and keeping a fixed accumulation time of $5$~ms in the reaction region. Setting the extraction RFQ to only transport $q=1+$ ions was found to fully suppress the formation of $^{208,212}$PbF$^{+}$ and $^{216}$PoF$^{+}$ ions, whereas small amounts of $^{224}$RaF$^{+}$ were identified under these conditions. The non observance of $^{208,212}$PbF$^{+}$ is in agreement with literature, where Pb monocations were found to not react with SF$_6$ gas \cite{cheng2009gas}.
Setting the extraction RFQ to transmit $q=2+$ ions, however, showed the production of $^{208,212}$PbF$^{+}$ and $^{216}$PoF$^{+}$ as well as $^{224}$RaF$^{+}$. 

These measurements show that Ra does react rather efficiently as both $q=1+$ and $q=2+$ ions with SF$_6$ forming $^{224}$RaF$^{+}$, whereas Pb and Po only react as doubly charged ions to form $^{216}$PoF$^{+}$ and $^{208,212}$PbF$^{+}$. Further, the observed SF$_5^{+}$ intensity roughly matches the sum of the $^{224}$Ra$^{2+}$, $^{216}$Po$^{2+}$ and $^{208,212}$Pb$^{2+}$ intensities without SF$_6$ gas. We conclude that SF$_5^{+}$ ions are formed in all of the observed fluorination reactions of doubly charged ions. 
However, individual experimental branching ratios could not be obtained in the present work. No reactions or molecules containing $^{208}$Tl$^{+}$ and $^{220}$Rn$^{+}$ were observed in the present study. 

In addition to the fluoridated species, $^{208,212}$PbOH$^{+}$ ions were observed as well when the extraction RFQ was set to transmit $q=2+$. 
The lead hydroxide seems to form via reactions with residual water vapor present in the reaction region, which was estimated to be on the order of $<1 \cdot 10^{-5}$~mbar, discussed below. 

\begin{figure}[hbt]
    \centering
    \includegraphics[width=0.8\linewidth]{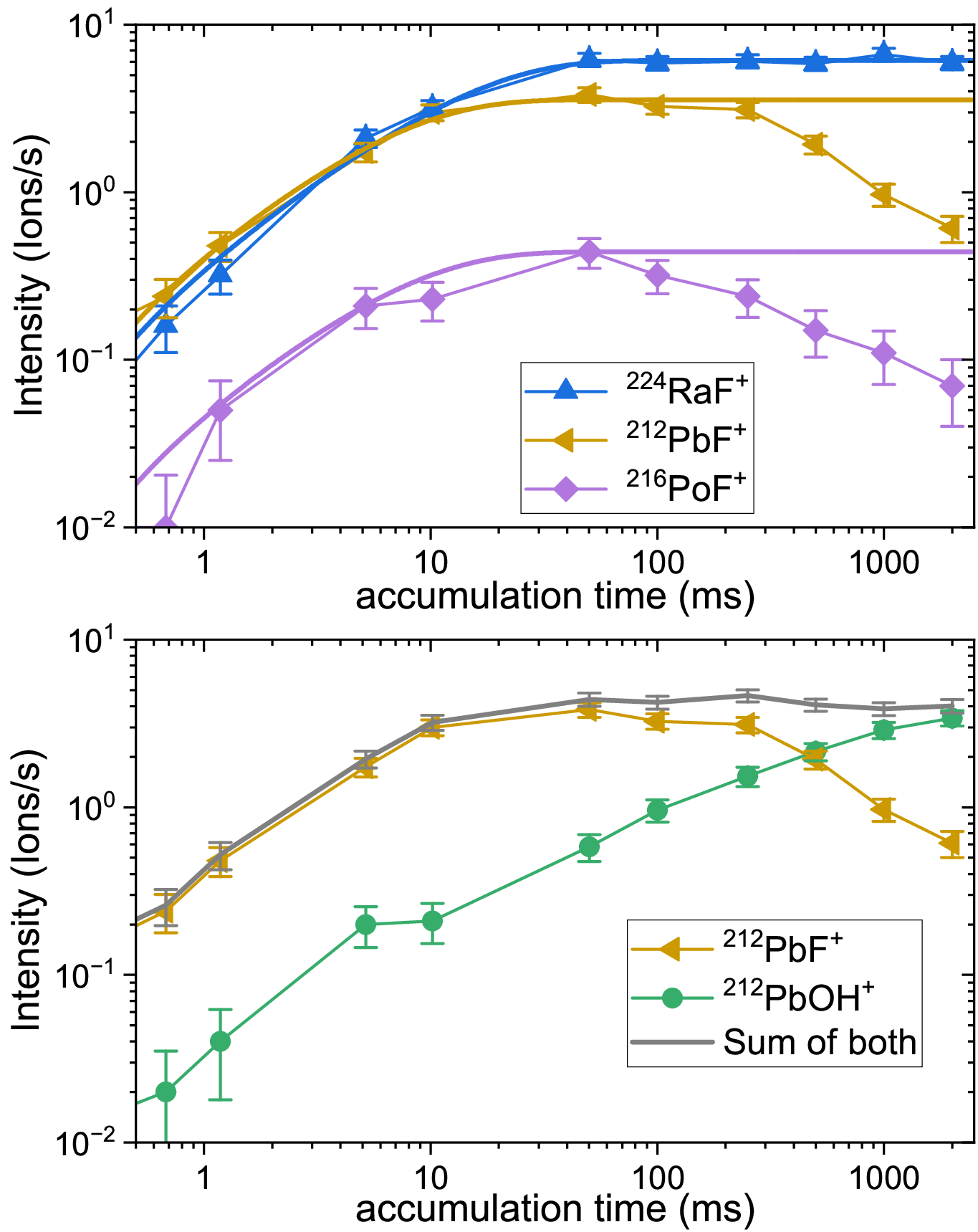}
    \caption{\textbf{Formation and secondary evolution of radioactive fluoride ions} (top) Detailed reaction study showing the formation of RaF$^+$, PoF$^+$ and PbF$^+$. Fit curves with a growth model are shown to guide the eye. (bottom) Formation of PbF$^+$ and PbOH$^+$ indicating the two steps reaction path $^{212}$Pb$^{2+}+$SF$_6\rightarrow^{212}$PbF$^{+}+$SF$_5^+$ and PbF$^+ +$H$_2$O$\rightarrow^{212}$PbOH$^{+}+$HF. }
    \label{fig:flourides}
\end{figure}

\subsection{Reaction rates of M$^{2+}$ ions with SF$_6$}\label{sec:react_rates}
After the initial charge state study of $q=1+$ and $q=2+$ reaction products, the build-up of molecular ions was studied. The growth functions were obtained by reducing the SF$_6$ pressure and monitoring the intensity of $^{224}$RaF$^{+}$, $^{216}$PoF$^{+}$ and $^{212}$PbF$^{+}$ for accumulation times ranging from $0.2$ to $2000$~ms (see Fig.~\ref{fig:flourides}, top). The measurement demonstrates the capability of the setup to measure reactions of several input and output channels simultaneously.   

After accumulation of approximately $50$~ms, the $^{224}$RaF$^{+}$ rate was saturated and the intensity stayed constant even after accumulation for $2000$~ms. In contrast, the intensities of both $^{216}$PoF$^{+}$ and $^{212}$PbF$^{+}$ grew with somewhat similar time constants but diminish after approximately $100$ and $200$~ms, respectively. In the case of $^{216}$PoF$^{+}$, the reduction can be well described by a first-order exponential decay, resulting in a half-life of $155(38)$~ms, in line with $^{216}$Po well-known half-life of $145(2)$~ms. 
However, in the case $^{212}$PbF$^{+}$, we find that its reduction is closely linked to the growth of $^{212}$PbOH$^{+}$. In Fig.~\ref{fig:flourides} (bottom) we show that the sum of $^{212}$PbF$^{+}$ and $^{212}$PbOH$^{+}$ remains constant after $^{212}$PbF$^{+}$ builds up.
Thus, a reaction of $^{212}$PbF$^{+}$ with residual amounts of H$_2$O seems to form $^{212}$PbOH$^{+}$. The reaction rate could not be quantified further since the concentration of residual water vapor in the reaction region was not determined in the present study. 

The measured SF$_6$ partial pressure and the individual decay and growth functions were used to calculate the fluoridation reaction rates of $^{224}$Ra$^{2+}$, $^{216}$Po$^{2+}$, and $^{212}$Pb$^{2+}$. Due to the low intensity of the $^{224}$Ra$^{+}$ beam, its reaction rate coefficients could only be estimated via the M$^+$ to M$^{2+}$ intensity ratios. As no reactions of $^{216}$Po$^{+}$ and $^{212}$Pb$^{+}$ ions were observed, only upper limits could be obtained.  In the present study, the overall uncertainty was dominated by the uncertainty in the SF$_6$ partial pressure (about $70 \%$) due to technical limitations of the pressure monitoring system, followed by the fit uncertainties of the growth and decay curves ($20$-$30 \%)$. These dominant contributions were combined in quadrature to obtain the rate coefficients as reported in Table~\ref{tab:coef}. 
Our measurements establish the fluorination kinetics at the proof-of-principle level, with the precision of the absolute bimolecular rate coefficients being limited mainly by the uncertainty in the SF$_6$ partial pressure. In the future, a dedicated residual gas analyzer (RGA) will be used to quantify the reaction-gas composition directly.

\begin{table}[h]
\normalsize
\caption{\normalsize \textbf{Effective bimolecular reaction rate constants $k$}, obtained for F-reactions of atomic M$^{+/2+}$ ions with neutral SF$_6$ gas,  measured at $P_{SF_6}=9(6)\cdot 10^{-5}$~mbar in Helium $P_{He}=9(4) \cdot 10^{-3}$~mbar at $301(5)$~K. For reactions not observed, only an upper limit is given.}
\begin{tabular}{l l l}
\toprule
M$^+$ ion & Product &  $k$     \\
 &  observed  &  (cm$^3$ molec.$^{-1}$ s$^{-1}$) \\
\midrule
 $^{224}$Ra$^{+}$  & $^{224}$RaF$^{+}$ &  $3(^{6}_{1.6}) \cdot 10^{-10}$ \\
 $^{224}$Ra$^{2+}$  & $^{224}$RaF$^{+}$ &  $7 (5) \cdot 10^{-10}$ \\
 $^{216}$Po$^{+}$  & not observed & $<10^{-12}$     \\
 $^{216}$Po$^{2+}$  & $^{216}$PoF$^{+}$ &  $10 (9) \cdot 10^{-10}$  \\
 $^{212}$Pb$^{+}$  & not observed&  $<10^{-12}$   \\
 $^{212}$Pb$^{2+}$  & $^{212}$PbF$^{+}$ &   $14 (10) \cdot 10^{-10}$  \\
\bottomrule
\end{tabular}
\label{tab:coef}
\end{table}

\subsection{Quantum chemical calculations}\label{sec:qc}

To help interpret the experiments, we used quantum‑chemistry calculations to estimate the energy demand or release pf the reactions studied. We first generated reference electronic structures with the restricted open‑shell Hartree–Fock (ROHF) method and then refined them using a coupled‑cluster approach that includes single and double excitation amplitudes with a perturbative treatment of triples [UCCSD(T)]; a widely used benchmark for small molecules.


Reaction energies were obtained by subtracting the total electronic energies of isolated reactants and products; vibrational and rotational contributions were not included. We report energies for (i) metal‑ion reactions with sulfur hexafluoride, M\textsuperscript{+} + SF\textsubscript{6} forming MF\textsuperscript{+} + SF\textsubscript{5} (M = Pb, Po, Ra), and (ii) subsequent reactions of the metal fluorides with water, MF\textsuperscript{+} + H\textsubscript{2}O forming MOH\textsuperscript{+} + HF. Numerical values and key intermediates are summarized in Table~\ref{tab:react_energ}. Further details and validation of the approach are provided in the Methods (Sections~\ref{sec:qc2} and \ref{sec:qc_val}).

\begin{figure}[!htb]
  \centering
  \includegraphics[width=0.9\linewidth]{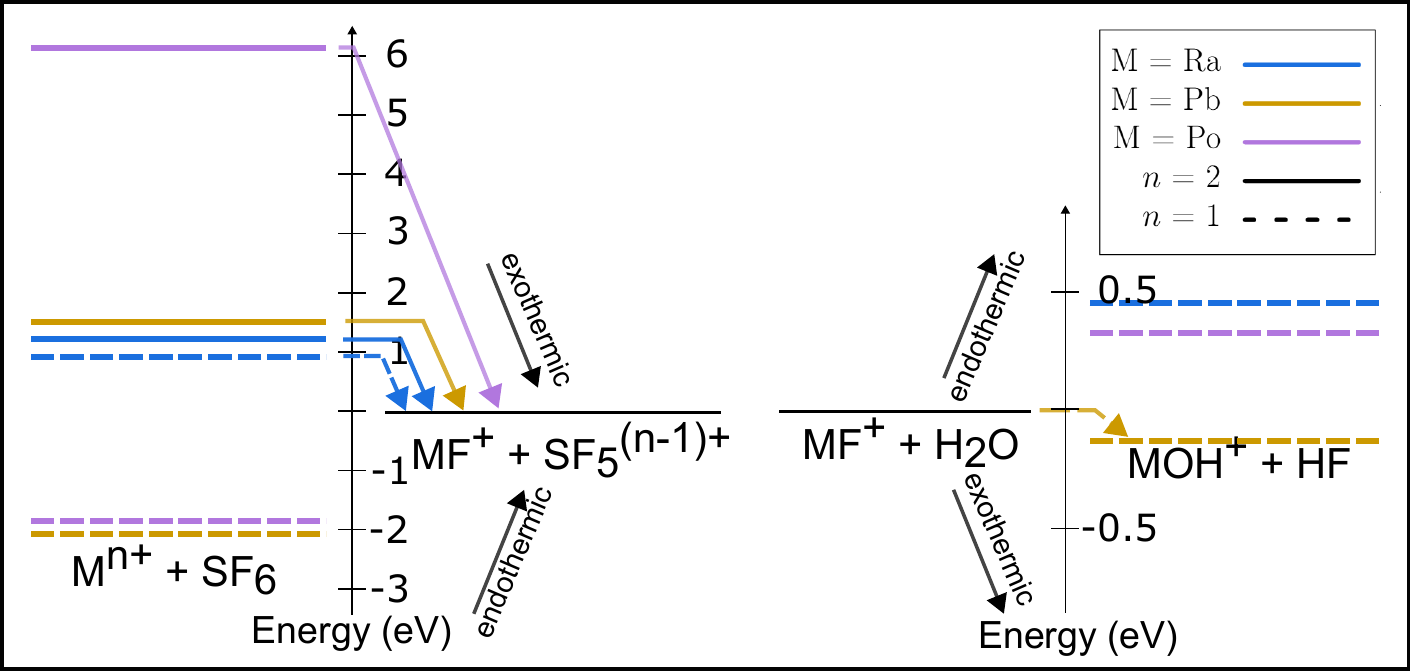}
  \caption{\textbf{Reaction-energy schematic for fluorination  and follow-up hydroxylation pathways.} Schematic reaction energy diagram showing the computed energy differences for the fluorination of M$ ^{+} $ (dashed lines) and M$ ^{2+} $ (solid lines) with SF$_6$ to MF$ ^+ $ (M $ = $ Ra, Pb, Po) , as well as the subsequent reaction MF$ ^+ $ + H$_2$O $ \rightarrow $ MOH$ ^+ $ + HF. Relative energies are given in eV, with the direction of the arrows  indicating endothermic or exothermic reaction pathways.  For the fluorination step shown on the left-hand side, the energies are referenced to the product channel, whereas for the hydroxylation pathways shown on the right-hand side they are referenced to the reactant channel. Only the mono-cation of Ra is expected with a negative reaction energy in the fluorination step, while all di-cations exhibit exothermic fluorination. The individual arrows indicate the specific experimentally observed reaction pathways, matching the theoretical calculations. } 
    \label{fig:react_energ}
\end{figure}



\begin{table*}[htb]
\centering
  \caption{\textbf{Reaction energies for the fluorination
  reaction}, energy differences (in eV) for M$^+$ and M$^{2+}$ (M=Pb, Po, Ra) with SF$_6$ computed on the level of
  RECP-ROHF-UCCSD(T), as well as energies for the follow-up reaction of MF$^+$ with SF$_6$.
  The stability of MF$^+$ towards hydroxylation with H$_2$O is
  estimated to assess the tendency to react with residual water vapor in the experimental setup.}
  \label{tab:react_energ}
  \begin{tabular}{
    l c l
    S[table-format=-2.4,round-mode=figures,round-precision=3]
    S[table-format=-2.4,round-mode=figures,round-precision=3]
    S[table-format=-2.4,round-mode=figures,round-precision=3]
  }
    \toprule
     & & & Ra & Pb & Po \\
    \midrule
    M$^{+}$  + SF$_6$             & $\rightarrow$ & MF$^+$ + SF$_5$   & -0.92 & 2.08   & 1.85   \\
    M$^{2+}$ + SF$_6$             & $\rightarrow$ & MF$^+$ + SF$_5^+$ & -1.22 & -1.51  & -6.13  \\
    M$^{2+}$ + SF$_5$ + F         & $\rightarrow$ & MF$^+$ + SF$_5^+$ & -6.04 & -6.14  & -10.76 \\
    M$^{2+}$ + SF$_5^+$ + F$^-$   & $\rightarrow$ & MF$^+$ + SF$_5^+$ & -12.43 & -12.60& -17.22 \\
    \midrule
    MF$^{+}$ + SF$_6$             & $\rightarrow$ & MF$_2$ + SF$_5^+$ & 3.90  & 4.74   & 2.37  \\
    MF$^{+}$ + SF$_5$ + F         & $\rightarrow$ & MF$_2$ + SF$_5^+$ & -0.91 & 0.12   & -2.26 \\
    MF$^{+}$ + SF$_5^+$ + F$^-$   & $\rightarrow$ & MF$_2$ + SF$_5^+$ & -7.30 & -6.34  & -8.720 \\
    \midrule
    MF$^{+}$ + H$_2$O             & $\rightarrow$ & MOH$^+$ + HF   &  0.45 &  -0.1322  & 0.323   \\
    \bottomrule
  \end{tabular}
\end{table*}

\section{Discussion}\label{sec12}

We compare our experimentally observed reaction pathways with the computed reaction energies. The quantum chemical calculations on the level of RECP-ROHF-UCCSD(T) (see Table~\ref{tab:react_energ}) predict the reaction energies for the individual paths M$^+$ + SF$_6$ $\rightarrow$ MF$^+$ + SF$_5$ all to be positive, with the exception of Ra$^+$ with \SI{-0.92}{eV}. Thus, the reactions of Pb$^+$ and Po$^+$ are predicted to be endothermic with energies \SI{2.08}{eV} and \SI{1.85}{eV} preventing Pb$^+$ and Po$^+$ from reacting efficiently with SF$_6$, respectively. This is in excellent agreement with observing experimentally only Ra$^{+}$ + SF$_6$ $\rightarrow$ RaF${^+}$ + SF$_5$ as the only M$^+$ reaction. 

The corresponding reaction energies for the M$^{2+}$ + SF$_6$ $\rightarrow$ MF$^+$ + SF$_5^+$ reaction paths are all exothermic with \SI{-1.22}{eV}, \SI{-1.51}{eV} and \SI{-6.13}{eV} for Ra$^{2+}$, Pb$^{2+}$ and Po$^{2+}$, respectively. The larger energy gain for the formation of PoF$^+$ can be attributed to the larger ionization energy (IE) of Po$^+$ (IE(Po$^+$) = \SI{19.4}{eV} \cite{hanni:2010} vs. IE(Pb$^+$) = \SI{15.0}{eV} \cite{finkelnburg:1955}). 
In Fig.~\ref{fig:react_energ} we illustrate the computed reaction energies. 
Products predicted here to be formed via an exothermic reaction channel were also observed in the experimental mass spectra.

Additionally, we consider subsequent reactions of MF$^+$ as potential loss channels, for example the formation of HF via the reaction MF$^+$ + H$_2$O $\rightarrow$ MOH$^+$ + HF. The calculations predict a slightly exothermic reaction with M$=$Pb with a reaction energy of \SI{-0.13}{eV} matching our accumulation time study, but endothermic reactions for Ra and Po.
Further, as without SF$_6$ no $^{212}$PbOH$^{+}$ formation was observed we conclude an intermediate reaction path with a two step reaction $^{212}$Pb$^{2+}+$SF$_6\rightarrow^{212}$PbF$^{+}+$SF$_5^+$ and $^{212}$PbF$^+ +$H$_2$O$\rightarrow^{212}$PbOH$^{+}+$HF to be most likely to occur. A similar reaction path with RaF$^+$, in contrast, is predicted as endothermic (reaction energy of \SI{0.45}{eV}), consistent with the plateau behavior observed in Fig.~\ref{fig:flourides}. Within the computational model, the energy for the reaction with PoF$^+$ is computed as \SI{0.323}{eV}, and, thus, endothermic as well. Furthermore, we predict only endothermic reactions to yield PoF$_2^+$, PoO, PoOF, or PoOF$^+$. 
Overall, the quantum chemical calculations predict all compounds MF$^+$ with M$=$Pb, Po, Ra to be thermodynamically stable towards dissociation by several eV, where the lowest dissociation channel would in all three cases be MF$^+$ $\rightarrow$ M$^+$ $+$ F. 
Together with the apparent absence of the relevant atomic fragment ions in experiment, this renders further chemical decomposition unlikely. 
Given $^{216}$Po's short half-life of $145(2)$~ms and non-observance of other Po molecules, radioactive decay seems to be the dominant source for the decline in the PoF$^+$ signal.


In Table~\ref{tab:compare}, the reaction rate coefficients of Ra$^{+/2+}$ with SF$_6$ are compared to those of other alkaline earth metals, all of which show very little variation around $6 \cdot 10^{-10}$~cm$^3$ molec.$^{-1}$ s$^{-1}$. Despite larger uncertainty, the rate coefficient of Ra$^{+}$ agrees well with Ca$^{+}$, Sr$^{+}$ and Ba$^{+}$. For Ra$^{2+}$, no direct comparison with previous literature was possible as reaction rates of Ca$^{2+}$, Sr$^{2+}$ and Ba$^{2+}$ are not known, but increased reactivity of higher charge states has been observed in the past, particularly for reactions of alkaline earth metal ions with O$_3$ \cite{10.1021/jp808077b,BOHME2022116674}.   

When comparing the outcomes of the respective reactions, a clear systematic trend can be seen. Whereas for Ca$^{+}$ + SF$_6$ a competition between the formation of CaF${^+}$ ($84\%$) and CaSF$_5^+$ ($16\%$) ions is observed in literature, for heavier alkaline earth metals this trend converges to almost a hundred percent MF$^{+}$ formation, as shown in Table~\ref{tab:compare}. In the present study, direct detection of potential RaSF$_5{^+}$ ions was not possible due to their high mass-over-charge. However, a limit on the branching ratio can be obtained by comparing the initial Ra$^{2+}$ rate with the RaF${^+}$ rate. A branching ratio limit of $90(10)\%$ was obtained for the  Ra$^{2+}$ + SF$_6$ $\rightarrow$ RaF${^+}$ + SF$_5^{+}$ reaction, in line with the overall trend from lighter alkaline earth metal ion reactions, but does not allow further conclusions.  

The reaction rate coefficients for Pb$^{2+}+$SF$_6\rightarrow$PbF$^{+}+$SF$_5^+$ and Po$^{2+}+$SF$_6\rightarrow$PoF$^{+}+$SF$_5^+$ could not be further compared to literature due to the lack of experimental data for both di-cations. Only the reaction of $^{212}$Pb$^{+}+$SF$_6$ was previously know to be endothermic \cite{cheng2009gas}, matching our findings.  

\begin{table}[!h]
\normalsize
\caption{\normalsize \textbf{Effective bimolecular reaction rate constants $k$ for F-reactions of atomic alkaline earth metal ions with neutral SF$_6$ gas in Helium.} Data for Ca$^+$, Sr$^+$ and Ba$^+$ are taken from \cite{cheng2009gas} and compared to Ra$^+$ and Ra$^{2+}$ from the present work. Due to the low statistics, no ratio could be obtained for Ra$^+$. The Langevin capture rate $k_\mathrm{c}$ was computed with the isotropic electric dipole polarizability volume $\alpha ($SF$_6)$ = \SI{6.54e-24}{cm^3} \cite{nelson1971,10.1021/ic061000o}, with the masses $m(\mathrm{SF}_6) = 146~\mathrm{u}$, $m(\mathrm{Ra}) = 224~\mathrm{u}$, $m(\mathrm{Ca}) = 44~\mathrm{u}$ and otherwise the standard atomic weights for Sr and Ba.  Note: For Ca, a mass of $44~\mathrm{u}$ was used to remain consistent with the literature value used for comparison, which was reported for $^{44}$Ca$^+$.
}
\begin{tabular}{l l r l l}
\toprule
M$^+$ & Product & Ratio & $k$  & $k_\mathrm{c}$    \\
 &   & ($\%$) & \multicolumn{2}{c}{(cm$^3$ molec.$^{-1}$ s$^{-1}$)}             \\
\midrule
 Ca$^{+}$  &  CaF$^{+}$  & $84$ & $6.3(1.9) \cdot 10^{-10}$   & $10.3\cdot10^{-10}$  \\
  &  CaSF$_5$$^{+}$ & $16$ & \\
  Sr$^{+}$  &  SrF$^{+}$  & $97$ & $5.7(1.7) \cdot 10^{-10}$  & $8.1\cdot10^{-10}$  \\
  &  SrSF$_5$$^{+}$  & $3$ & \\
   Ba$^{+}$  &  BaF$^{+}$  & $99$ & $6.1(1.8) \cdot 10^{-10}$ & $7.1\cdot10^{-10}$  \\
  &  BaSF$_5$$^{+}$  & $1$ & \\
 \midrule
 Ra$^{+}$  &  RaF$^{+}$ &   & $3(^{6}_{1.4}) \cdot 10^{-10}$  & $6.4\cdot10^{-10}$  \\
Ra$^{2+}$  &  RaF$^{+}$ & $90(10)$  & $7(5) \cdot 10^{-10}$   & $13\phantom{.}\cdot10^{-10}$  \\
\bottomrule
\end{tabular}
\label{tab:compare}
\end{table}

We have demonstrated a compact, accelerator-independent method to produce and study short-lived, heavy radioactive molecular ions at trace levels by combining alpha-recoil harvesting in a high-purity stopping cell with fast, selective ion–molecule reactions in a modular RFQ beamline and MR-TOF mass spectrometry. We generated and unambiguously identified RaF$^+$, PoF$^+$, and PbF$^+$, quantified reaction kinetics, and established a clear, element- and charge-state–dependent reactivity pattern. The observed doubly-charged ions react efficiently with SF$_6$ to form monofluorides, whereas singly charged channels are strongly suppressed for Pb and Po but remain possible for Ra. For Ra$^{2+}$ + SF$_6$, we observe near-unity conversion to RaF$^+$ on millisecond timescales. High-resolution mass spectrometry reveals subsequent hydroxylation of PbF$^+$ to PbOH$^+$ in the presence of residual water, while RaF$^+$ remains inert under identical conditions. Quantum chemical computations on the RECP-ROHF-UCCSD(T) level explain these trends: formation of MF$^+$ from M$^{2+}$ + SF$_6$ is exothermic for Ra, Pb, and Po, whereas M$^{+}$ + SF$_6$ is endothermic for Pb and Po but not for Ra, in agreement with experiment.

The production and detection of $^{216}$PoF$^+$
from $^{216}$Po$^{2+}$ - despite the short half-life of $145$~ms - provides a direct proof-of-concept  that the full chain of processes (alpha-recoil emission, thermalization, extraction, guided transport, reaction, and mass-resolved detection) proceeds on short timescales. The measurable buildup and subsequent decay of the $^{216}$PoF$^+$ signal confirm that our scheme accesses short-lived species well below the one-second regime, overcoming a key bottleneck for decentralized studies of exotic heavy nuclei and their molecules.

This scheme addresses central limitations that have hindered decentralized, university-scale studies of molecules bearing post-lead elements: it operates with sub-microgram/nanogram samples at kBq activities, proceeds on millisecond timescales compatible with short half-lives, and provides high chemical specificity via mass-resolved detection. The demonstrated efficiency of RaF$^+$ production from Ra$^{2+}$ + SF$_6$, together with the observed chemical stability of RaF$^+$, establishes a practical pathway to supply Ra-based molecular ions for precision measurements, including EDM searches. More broadly, the method enables systematic gas-phase chemistry of heavy elements beyond lead and supports applications in radiochemistry and medical radionuclide development, where isotope scarcity and short lifetimes have impeded progress.
Scaling of the present harvesting scheme is governed primarily by the recoil source. Scaling is not linear with source thickness, because only recoils produced sufficiently close to the source surface can escape into the stopping gas \cite{reed2026}. Source activities in the range of about $ 300 $~kBq have be achieved by increasing the source area rather than the source thickness \cite{v.d.Wense2015}, whereas activities approaching the $10-100$~MBq level will in general require a correspondingly larger source surface and hence a dedicated larger setup. 

Looking ahead, a dedicated Radium-Fluoride Ion Catcher Instrument (RAFICI) will extend this platform by: diversifying the accessible set of precursor–product chains beyond $^{208}$Pb, increasing throughput and stability for targeted Ra isotopes relevant to EDM experiments, exploring additional reagents (e.g., NF$_3$, CF$_4$) and routes to polyatomic species (e.g., RaOH$^+$, RaOCH$_3^+$, RaNH$_3^+$).
At the FRS Ion Catcher, the dominant systematic uncertainty in the rate coefficients will be reduced in future measurements by an in-situ gas pressure and composition diagnostics (residual gas analysis and absolute pressure measurement), refined accumulation-time scheme in the reaction region, and benchmarks against reaction theory and known stable species. 

In summary, alpha-recoil harvesting combined with ion–molecule chemistry and high-resolution mass analysis provides an efficient, portable, and generalizable route to heavy radioactive molecular ions. 
 It complements existing trapped-ion and cryogenic neutral-molecule platforms by providing a route to controlled, charge-state-selective reaction studies and broadband mass-resolved identification of radioactive molecular ions.
The successful formation and observation of $^{216}$PoF$^+$ underscores the capability to access short-lived isotopes, opening a path to systematic, decentralized studies of molecules containing post-lead elements, which underpin precision tests of fundamental symmetries and advance heavy-element chemistry and applications.

\section{Methods}\label{sec14}
\subsection{Experimental scheme}\label{exp}
The experiment was performed at the FRS Ion Catcher facility, as shown schematically in Fig.~\ref{fig:setup}~top. 
An open (non-covered) $^{228}$Th source (Eckert and Ziegler, about $7$~kBq) was installed inside the DC cage of the CSC. The cell was filled with 55~mbar ultra pure He gas and cooled to 85~K. Under these conditions the alpha decay recoils were stopped within 2~mm from the source surface. Once stopped and thermalized in the gas, a combination of DC and RF electric fields was used to extract the ions. 
A voltage gradient of 16~V/cm was applied to guide the ions along the 0.5~m long body of the CSC towards the exit side. Here, an RF carpet provided a pseudo-potential wall preventing the ions from hitting the surface. The RF carpet was driven with a 60~V$_{pp}$ $6.1$~MHz RF signal, which was superimposed on a static DC gradient of 5~V/cm, focusing the ions towards the center,  resulting in extraction times in the order of $60$~ms. Once guided to the central extraction nozzle, the gas flow guided the ions into the differentially pumped radiofrequency quadrupole (RFQ) beamline. 
The ions are transported along the RFQ RFQ beam line via a small voltage gradient of approximately 0.5~V/cm
The beam line, typically operated at a residual gas pressure of $P_{He}=9(4) \cdot 10^{-3}$~mbar He, takes care of multiple tasks. First, its extraction RFQ can be operated as a low-resolution mass filter, which allows selecting different charge states of an isotope. Second, it is equipped with multiple detectors to identify and quantify extracted ions via single-ion counting and alpha spectroscopy. Third, it allows accumulation, buffer gas cooling, and re-bunching of the extracted continuous ion beam. 
Further, small amounts of reaction gas, in our case SF$_6$ with $P_{SF_6}=9(6)\cdot 10^{-5}$~mbar (SF$_6$ gas 5.0, Linde), can be mixed with the He buffer gas in the reaction chamber of the beamline. Ions can be trapped and stored for a certain amount of time, and reaction products as well as unreacted ions can be injected into a Multiple-Reflection Time-Of-Flight Mass-Spectrometer (MR-TOF-MS). The latter, is a high-resolution mass spectrometer where ions are stored between two electrostatic ion mirrors. Path length of up to a~km can be reached in the compact device, resulting in mass resolving power of up to $10^6$ .

\subsection{Yield and sample purity}

 In this source-based scheme, the nuclides accessible are those populated in $ \alpha $-decay, whereas nuclei produced in $ \beta $-decay steps are typically not released from the source surface with comparable efficiency. In our case relevant recoil-accessible decay sequence is $ ~^{228}\mathrm{Th} \xrightarrow{\alpha}  ~^{224}\mathrm{Ra} \xrightarrow{\alpha} ~^{220}\mathrm{Rn} \xrightarrow{\alpha} ~^{216}\mathrm{Po} \xrightarrow{\alpha} ~^{212}\mathrm{Pb} \xrightarrow{\beta^-} ~^{212}\mathrm{Bi} $, followed by the branches $ ~^{212}\mathrm{Bi} \xrightarrow{\alpha} ~^{208}\mathrm{Tl} $ and $ ~^{212}\mathrm{Bi} \xrightarrow{\beta^-} ~^{212}\mathrm{Po} \xrightarrow{\alpha} ~^{208}\mathrm{Pb} $, a graphical representation of the decay scheme can be found in \cite{miskun2019}. 
The installed source had a total activity of about $ 7 $~kBq. 
Under the present operating conditions, extraction from the stopping cell occurs on the $60$~ms timescale, such that the recoil daughters studied here are removed from the cell well within their respective half-lives. 

 The main contributions to the observed ion yield are the release efficiency from the source, the stopping and extraction efficiency from the stopping cell, and the transport efficiency downstream of the reaction region and into the MR-TOF-MS and final detection, together close to a percent in the present experiment.

Thus $^{224}$Ra$^{+/2+}$, $^{216}$Po$^{+/2+}$ and $^{212}$Pb$^{+/2+}$ ions were each detected with about $5-15$ ions per second at the MR-TOF-MS. 

\begin{figure}[htb]
    \centering
    \includegraphics[width=0.8\linewidth]{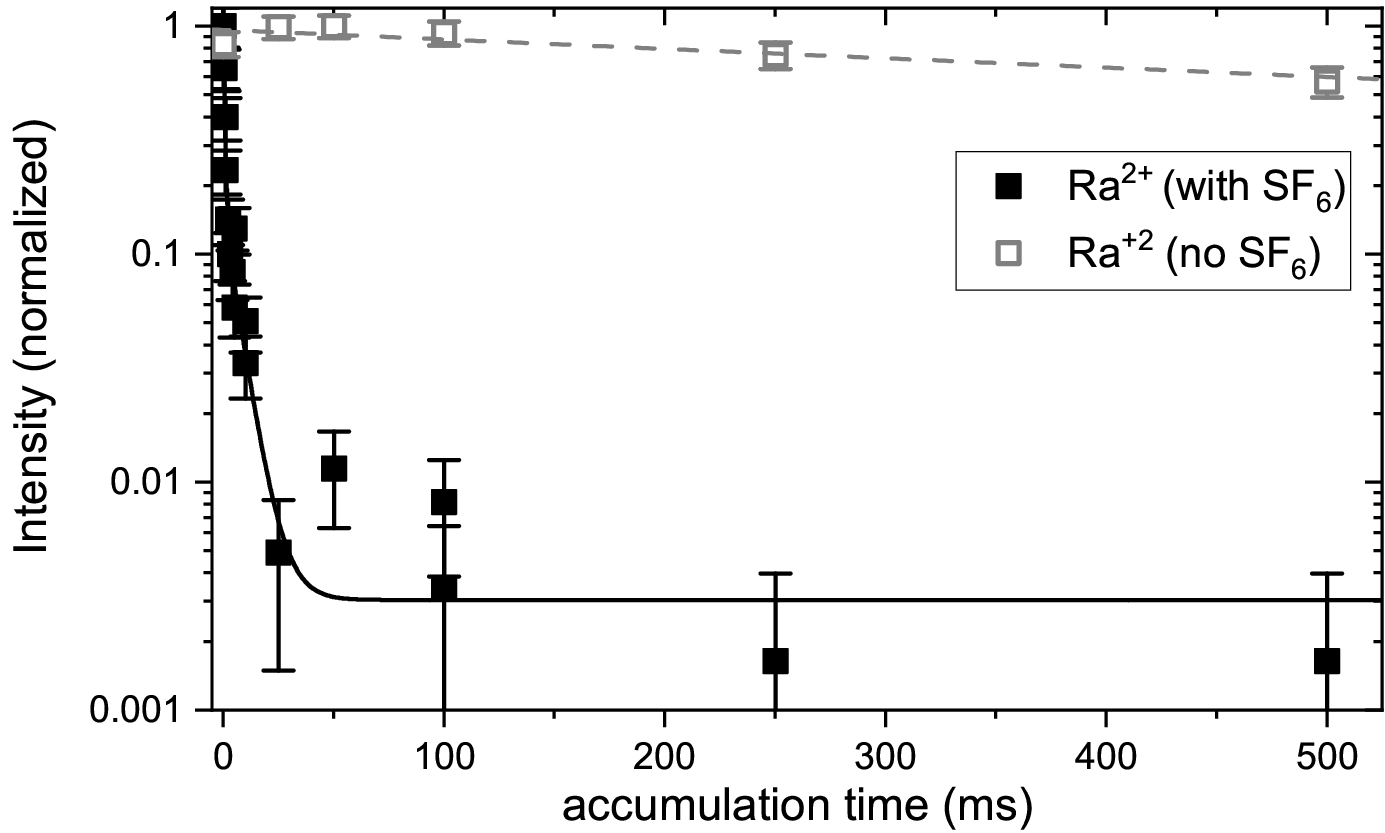}
    \caption{\textbf{Reference accumulation study of $^{224}$Ra$^{2+}$}, showing the survival of $^{224}$Ra$^{2+}$ ions with and without SF$_6$ applied to the transport RFQ section. Fit line to guide the eye. The intensity of $^{224}$Ra$^{2+}$ at long accumulation times is estimated as an upper limit based on the non-observance of events.}
    \label{fig:Radium_decay2}
\end{figure}

In order to validate the purity of the helium buffer gas in the reaction region and to rule out reactions with unwanted impurities, the ions were stored in the reaction region without SF$_6$ gas added. Ra$^{2+}$ ions could be stored for up to 0.5~s with only minimal losses, as shown in Fig.~\ref{fig:Radium_decay2}. Similar storage times are assumed for the other elements, but were not validated further.   

The SF$_6$ gas pressure  (obtained from a hot cathode gauge, MKS) was calibrated against the setpoint of a motorized needle valve (Pfeiffer, EVR116) by first reducing the He residual gas pressure to a minimum and then changing the SF$_6$ flow. A linear behavior was seen, the flow was then set to the lowest and 2nd lowest reproducible setpoints,  resulting in an SF$_6$ partial pressure of $P_{SF_6}=9(6)\cdot 10^{-5}$~mbar.  

 For the reaction-rate analysis, the ion-population evolution was treated in the effective first-order limit, and the precursor-ion decay and product-ion growth curves were fitted by least-squares minimization with single-exponential functions of the form 
\begin{eqnarray} 
N(t)=N_0 \exp(-t/\tau) 
\end{eqnarray}
and 
\begin{eqnarray} 
N(t)=N_{\infty}\left(1-\exp(-t/\tau)\right),
\end{eqnarray}
respectively; the corresponding bimolecular rate coefficients were then obtained from the fitted time constants using the estimated SF$_6$ number density. 

\subsection{Mass spectrometry}
The time-of-flight mass spectrometry relies on the relation between the mass $m$ and charge $q$ of an ion and the time-of-flight $t_\mathrm{tof}$ required to travel a certain flight path $l$ through an electric potential $U(l)$: 
\begin{eqnarray}
\frac{m}{q}= \frac{2U(l)}{l^2} t_\mathrm{tof}^{2} 
\label{eq:mass-tof3}
\end{eqnarray}
However, due to signal propagation times and other electronic delays, the real time of flight $t_\mathrm{tof}$ is often different from the measured time of flight $t_\mathrm{exp}$, requiring a modified calibration function that is: 
\begin{eqnarray}
\frac{m}{q} =a (t_\mathrm{exp} - t_{0})^{2} \label{eq:mass-tof}
\end{eqnarray}
with $a$ being a device-specific calibration parameter, 
which together with the offset $t_{0}$ can be determined from calibration measurements using ions of known mass-to-charge ratios. 

The resolving power of such a time-of-flight mass spectrometer can be increased by storing ions for multiple reflections between two electrostatic isochronous ion mirrors, expanding the effective flight path. 
At long flight paths, a resolving power on the order of $\frac{m}{\Delta m}$ of $ \sim 10^{5}$ can be achieved. 

\begin{figure}[htb!]
    \centering
     \includegraphics[width=0.7\linewidth]{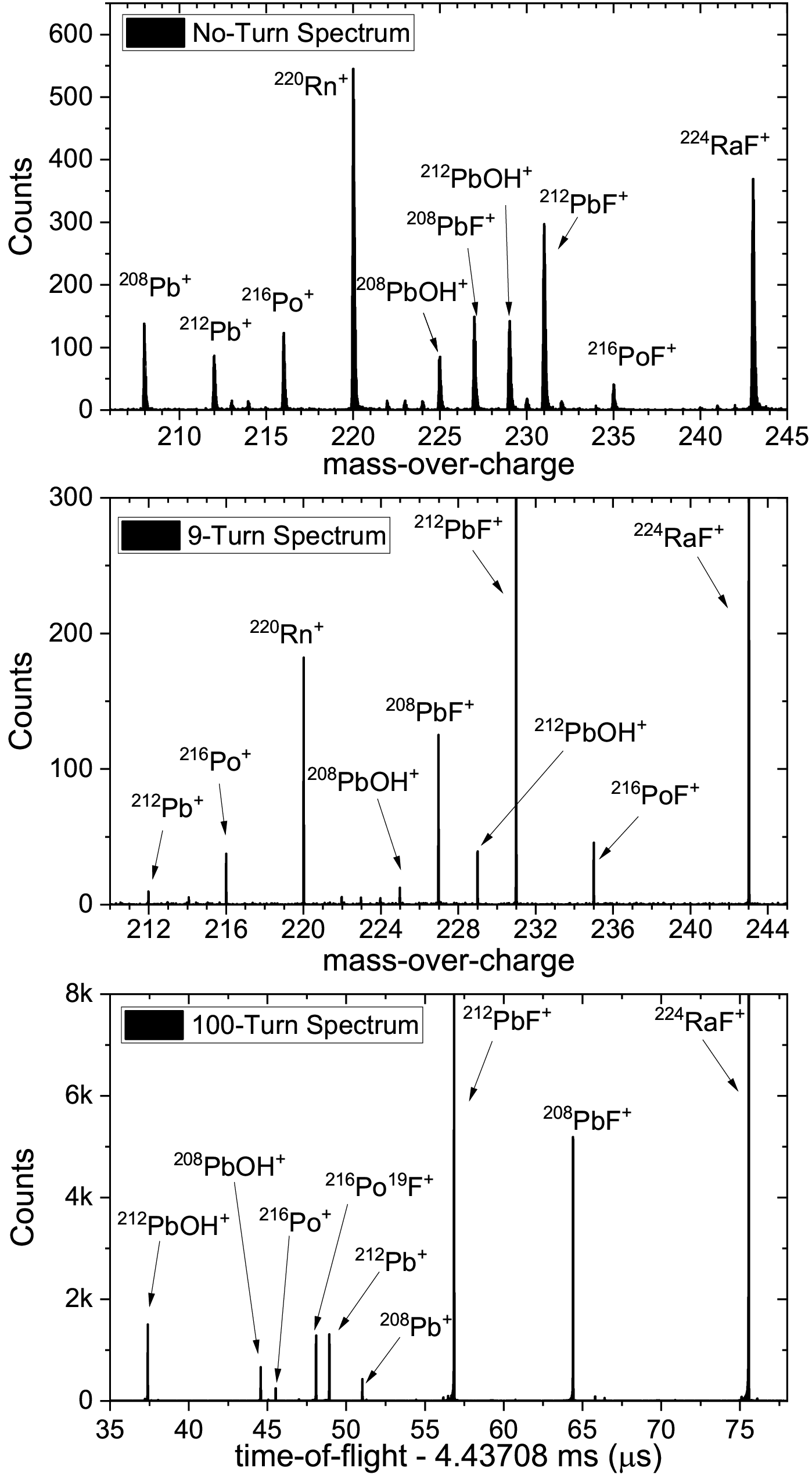}
    \caption{\textbf{Representative MR-TOF-MS mass spectra} Three example mass spectra taken with the MR-TOF-MS during this experiment in a) No-Turn mode  (resolving power $2\ 000$), b) 9-Turn mode (resolving power $20\ 000$) and c) 100-Turn mode (resolving power $100\ 000$). Note, due to the closed flight path inside the MR-TOF-MS ions with different mass numbers undergo a different number of turns, which is account for using eq.~\ref{eq:calib}.}
    \label{fig:mass_spec_overview}
\end{figure}

However, if one has a broad initial mass distribution and a high number of turns, lighter ions can overtake heavier ones. Thus, the mass spectrum can become ambiguous, as ions undergo a different number of turns inside the mass analyzer for a given TOF. In this case, equation \ref{eq:mass-tof3} needs to be modified to account for the different flight paths the ions have undergone. Using the $s_\mathrm{offset} $ for the length of the flight path from the start to the final detector without any turns and $s_{\mathrm{turn}}$ for the length of one turn, the total flight path $l$ can then be written as $l=s_\mathrm{offset} +N \cdot s_{\mathrm{turn}} $, yielding a calibration function of
\begin{eqnarray}
\frac{m}{q} =  \frac{2U(l) }{(s_\mathrm{offset} +N \cdot s_{\mathrm{Turn}})^{2}} (t_\mathrm{exp}-t_{0})^{2}. \label{eq:mass-tof2}
\end{eqnarray}
This expression can be simplified to
\begin{eqnarray}
\frac{m}{q} = \frac{c(t_\mathrm{exp}-t_{0})^{2}}{(1+N \cdot b)^{2}},  \label{eq:calib}
\end{eqnarray}   
with $b=s_\mathrm{turn}/s_\mathrm{offset}$ and $c=2U(l)/s_\mathrm{offset}^2$.
Thus, a turn-independent calibration, using three known calibration species with at least two different numbers of turns $N$ can determine $b$, $c$, and $t_{0}$ \cite{PhysRevC.99.064313}. In the present case, $^{212}$Pb$^{+}$, $^{216}$Po$^{+}$,  $^{220}$Rn$^{+}$, and $^{224}$Ra$^{+}$ were used for the calibration prior to performing the molecule studies using SF$_6$ gas. In the high resolution settings then used, $^{224}$Ra$^{+}$ ions were not transmitted to the time-of-flight detector.      

In the present work, the MR-TOF-MS was operated in a $0$-, $9$-, and $100$-turn modes. Example mass spectra are shown in Fig.~\ref{fig:mass_spec_overview}. For the mass range $206-245$ in the $100$-turn mode, a turn-independent calibration function was used.  Because the measurement is based on single-ion counting and all ion species of interest were well separated in time-of-flight, the relative intensities used in the gas-phase chemistry analysis were obtained by integrating the number of detected events within narrow time-of-flight windows around each peak, while gaussian peak-shape fits were used for peak validation and mass determination. All peaks were confirmed based on their known mass-over-charge with a precision better than $1$~ppm.


\subsection{Quantum chemical calculations}\label{sec:qc2}
Calculations 
were performed with the quantum chemistry program package \textsc{Molpro}
\cite{molpro2020} on the level of restricted open-shell Hartree--Fock (ROHF). These calculations were used as a reference for subsequent unrestricted coupled-cluster computations
with iterative single and double excitations, including perturbative triple excitation amplitudes [UCCSD(T)]. 
Relativistic effects
were incorporated in this non-relativistic framework by using scalar relativistic effective core potentials (RECP) on
the metal center. The wavefunction on the level of ROHF was optimized in a self-consistent-field manner until
the change in energy between two consecutive cycles was below $10^{-13}~E_\mathrm{h}$. Bond lengths were optimized until the cartesian gradient norm was below $10^{-5}~E_\mathrm{h}/a_0$.

On the hydrogen, carbon, nitrogen, oxygen and fluorine center augmented correlation consistent basis sets of
triple-$\zeta$ quality were employed (aug-cc-pVTZ) \cite{dunning:1989,kendall:1992}. For sulfur a basis set of the same family with double-$\zeta$ quality (aug-cc-pVDZ) \cite{dunning2001} was employed. For radium a 78 electrons in core ECP \cite{lim:2006} and for lead \cite{metz:2000} and
polonium \cite{peterson:2003} a 60 electrons in core ECP were used. The ECPs were complemented by an atomic natural orbital valence basis set of triple-$\zeta$ quality. For further assessing the quality of the ECP on Po, we computed the first and second ionization energy and electron affinity of Po with a basis of quadruple-$\zeta$ and quintuple-$\zeta$ quality as well. 

In the UCCSD(T) computations, electrons in 1s orbitals of O and F were omitted from the correlation treatment (frozen core approximation). For Pb and Po the 5s, 5p and 6d orbitals and for S the 1s, 2s, 2p orbitals are included in the frozen core. In computations involving Ra electrons in all orbitals at Ra were correlated.
The structure of SF$_6^+$ was taken from the UCCSD(T)-F12b results in Ref.~\cite{simpson2025}. From the same reference the UCCSD(T) structure employing a basis set of double-$\zeta$ quality is taken for SF$_5$. Here, we optimized the bond lengths of SF$_5^+$ with $D_{3h}$ symmetry constraints on the level of UCCSD(T) with all electrons correlated, employing the aug-cc-pVDZ basis set on the F center. We highlight that we did not include zero-point energies, neither did we account for basis set superposition error nor included spin orbit coupling contributions.

\subsection{Validation of calculations}\label{sec:qc_val}
The ECPs used were assessed on the ionization energies of the atoms and, in the case of
Polonium, on an estimate of the electron affinity.
A comparison of various ionization energies, electron affinities, and dissociation energies is provided in the in Supplementary Section, Supplementary Table~1 and Supplementary Table~2, in order to assess the limits of the  methodology used herein.
Radium and lead are experimentally well-investigated systems. The former has a comparatively simple
electronic structure for the neutral, mono, and doubly ionized system,
which are well describable in the ECP-ROHF-UCCSD(T) approach, i.~e. with a
single determinant. Computations of the corresponding first (\SI{5.23}{eV} vs.
\SI{5.28}{eV} \cite{armstrong:1980}) and second (\SI{10.1}{eV} vs.
\SI{10.1}{eV} \cite{dammalapati:2016}) ionization energy of Ra and also the
ionization energy of RaF (\SI{4.95}{eV} vs. \SI{4.97}{eV} \cite{wilkins2024ionization})
agree with the corresponding experimental data from literature. The lead ground
state is more difficult to describe in a single determinant approach due to its
$^3P_0$ electronic state. The first ionization energy calculated differs by
approximately \SI{0.55}{eV} (\SI{6.87}{eV} vs.  \SI{7.42}{eV}
\cite{dembczynski:1994}) compared to experimental data from literature. It is
mentioned that the data agrees with the computed first ionization energy
obtained as a test for the ECP in Ref.~\cite{metz:2000}. In Ref.~\cite{metz:2000}
a correction to the experimental data by \SI{0.31}{eV} for further missing
spin-orbit interactions is used for the first ionization energy, which we can
take here for a rough error estimation. The deviation of the here computed
second ionization energy compared to experimental literature values is
\SI{1.6}{eV} (\SI{13.4}{eV} vs.  \SI{15.0}{eV} \cite{hanni:2010}) exhibiting
the lower accuracy and limits of the used methodology for higher ionized
states.

Polonium exhibits a similar case, where the ground state, and also the first
and second ionized state, has a multi-reference character. The electron affinity
of \SI{1.9}{eV} used to test the ECP in Ref.~\cite{peterson:2003} is perfectly
reproduced here. Current literature on the electron affinity of Polonium
predicts a value of \SI{1.4}{eV} \cite{borschevsky:2015} including Gaunt and
Breit terms. The first ionization energy computed here has a similar deviation
as in the case of Pb, compared to the literature experimental value
(\SI{8.08}{eV} vs. \SI{8.42}{eV} \cite{fink:2019}). Higher cardinality of
quintuple-$\zeta$ quality with the same ECP increases the here computed value
to \SI{8.22}{eV}. The deviation of the second ionization energy compared to
experimental literature values is on the order of IE(Pb$^+$) (\SI{17.9}{eV} vs.
\SI{19.4}{eV} \cite{finkelnburg:1955}). 


\backmatter


\bmhead{Data availability}
The data generated in this study are provided in the Supplementary Information/Source Data file.


\bmhead{Acknowledgements}
We would like to thank Peter B. Armentrout for intial discussions. The work was supported by the Royal Society no.\ RGS-R2-222093 (12886009), by the Edinburgh nuclear physics consolidated grant, UKRI STFC under grant agreement no.\ ST/Y000293/1, by the German Federal Ministry of Research, Technology and Space (BMFTR)
under contracts no. 05P21RGFN1 and 05P24RG4, by HGS-HIRe, and by Justus-Liebig-Universit{\"a}t Gie{\ss}en and GSI under the JLU-GSI
strategic Helmholtz partnership agreement. 
The authors gratefully acknowledge computing time made available to them on the high-performance computers Goethe-HLR at the NHR Centers NHR S{\"u}d-West. These Centers are jointly supported by the Federal Ministry of Education and Research and the state governments participating in the NHR (\url{http://www.nhr-verein.de/unsere-partner}). Financial support by the Deutsche Forschungsgemeinschaft (DFG, German Research Foundation) with the project number 445296313 is gratefully acknowledged.

\bmhead{Funding Statement}
Royal Society no.\ RGS-R2-222093 (12886009) \newline
UKRI STFC no.\ ST/Y000293/1 \newline 
German Federal Ministry of Research, Technology and Space (BMFTR) no. 05P21RGFN1 and 05P24RG4, \newline 
Deutsche Forschungsgemeinschaft (DFG, German Research Foundation) no.\ 445296313.

\bmhead{Author contributions}
MPR and TD conceived the project and developed the methodology. 
MPR, KM, MN, TD, DA, RB, AB, TF-D, ZG, SG, GK-K, CM, WRP, MS, NT, JY, JZ, MF and CZ carried out the investigation using analysis software / codes developed by JB and MS. KM and GK-K curated the data. MPR, KM, MN, CZ, TD and DJM performed the formal analysis. DJM and AZ validated the results.
The project was supervised by MPR, TD, RB, DJM, CS, NK-N and WRP. Funding was acquired by MPR, TD, RB, DJM, CS, NK-N and WRP. 
All authors contributed to writing, reviewing and editing the manuscript.

\bmhead{Conflict of interest/Competing interests}
The authors declare no competing financial or non-financial interests.


\end{document}